\documentclass[]{pasj02} 
\usepackage[switch,mathlines]{lineno} 
\usepackage{graphicx}
\usepackage{url}
\usepackage{xcolor}

\usepackage[normalem]{ulem}
\jyear{2026}
\Received{2026/3/29}
\Accepted{2026/9/20}
\Published{2026/11/13}

\begin{document} 
\title{XRISM/Xtend detection of solar wind charge exchange induced by multiple coronal mass ejections in 2024 August}

\author{
 Daijiro \textsc{Kanno},\altaffilmark{1}\altemailmark\email{kanno-daijiro@ed.tmu.ac.jp} 
 Daiki \textsc{Ishi},\altaffilmark{2}
 Yuichiro \textsc{Ezoe},\altaffilmark{1}
 Masaki \textsc{Numazawa},\altaffilmark{1}
 Kumi \textsc{Ishikawa},\altaffilmark{1}
 Shogo B \textsc{Kobayashi},\altaffilmark{3}
 Satoru \textsc{Katsuda},\altaffilmark{4}
 Liyi \textsc{Gu}\altaffilmark{3,5,6,7}
}
\altaffiltext{1}{Department of Physics, Tokyo Metropolitan University, 1-1 Minami-Osawa, Hachioji, Tokyo 192-0397, Japan}
\altaffiltext{2}{Institute of Space and Astronautical Science, Japan Aerospace Exploration Agency, 3-1-1 Yoshinodai, Chuo-ku, Sagamihara, Kanagawa 252-5210, Japan}
\altaffiltext{3}{Department of Physics, Tokyo University of Science, 1-3 Kagurazaka, Shinjuku-ku, Tokyo 162-8601, Japan}
\altaffiltext{4}{Graduate School of Science and Engineering, Saitama University, 255 Shimo-Ohkubo, Sakura, Saitama 338-8570, Japan }
\altaffiltext{5}{SRON Space Research Organisation Netherlands, Niels Bohrweg 4, 2333 CA Leiden, The Netherlands}
\altaffiltext{6}{Leiden Observatory, Leiden University, PO Box 9513, 2300 RA Leiden, The Netherlands}
\altaffiltext{7}{RIKEN High Energy Astrophysics Laboratory, 2-1 Hirosawa, Wako, Saitama 351-0198, Japan}



\maketitle

\begin{abstract}
Solar wind charge exchange (SWCX) emission is often a major component of the soft X-ray background, but it also provides a remote diagnostic of the composition and ionization state of solar wind plasma. We present an analysis of a SWCX event observed with XRISM/Xtend during 2024 August 9--13. The event occurred during intense solar activity with multiple flares, including an X1.3-class solar flare, and associated coronal mass ejections (CMEs). We found a clear enhancement in the 0.3--3.0~keV count rate that occurred nearly simultaneously with an increase in the solar wind proton flux measured by ACE. We fitted the spectrum during the enhancement period with a set of Gaussian emission lines. In addition to the commonly detected C, N, and O lines, we detected Ne, Mg, Si, and Fe lines at $>15\sigma$ significance. We extracted two characteristic periods within the enhanced interval based on the CME arrival timeline. The O$_{\rm VIII}$/Si$_{\rm XIII}$ line ratio changed significantly between the two periods, suggesting contributions from multiple CME components. We then applied the AtomDB charge exchange model with a multi-temperature setup. The estimated ionization temperatures for Ne, Mg, Si, and Fe increased with the peak soft X-ray flux of the associated solar flares. This trend suggests that coronal heating becomes more efficient for more intense flares. Meanwhile, the ionization temperatures were lower than typical flare electron temperatures, implying that CME plasma may have been released into interplanetary space under non-equilibrium ionization conditions. We also found an indication of an inverse-FIP pattern, with reduced relative abundances of Si and S. These results demonstrate that XRISM/Xtend can diagnose the coronal plasma through SWCX emission, including heavy ions such as Ne, Mg, and Si, which are difficult to measure with current in-situ instruments (e.g., ACE satellite).
\end{abstract}


\noindent {\small This is a pre-copyedited, author-produced version of an article accepted for publication in ``Publications of the Astronomical Society of Japan'' following peer review. The version of record (PASJ, 2026, Vol.~78, Issue~6) is available online at \url{https://doi.org/10.1093/pasj/psag143}. \par}

\section{Introduction}
The Sun is surrounded by a hot atmosphere, the corona, with temperatures reaching $\sim$1~MK. The mechanism responsible for heating the corona to such high temperatures remains poorly understood, a puzzle known as the coronal heating problem (\cite{Klimchuk06}). Coronal mass ejections (CMEs), often associated with solar flares, provide key insights into this issue by ejecting large quantities of highly ionized plasma into interplanetary space. During this process, the ionization states of the plasma become ``frozen-in'' within the corona (\cite{Geiss+95}). Consequently, the ionization state remains largely unchanged as the plasma travels from the Sun to near-Earth space. This preserves information regarding the physical conditions in the corona. Thus, measurements of the ionization state of CME plasma near Earth allow for the indirect constraint of coronal plasma conditions (\cite{Reinard05}). However, since an instrumental anomaly in 2011\footnote{\sloppy\url{https://izw1.caltech.edu/ACE/ASC/level2/ss2_l2desc.html}}, the Solar Wind Ion Composition Spectrometer on board the Advanced Composition Explorer (ACE) has been unable to provide composition data for heavy elements such as Ne, Mg, and Si.

Solar wind charge exchange (SWCX) offers an alternative approach to investigating the composition and ionization state of the solar wind. In this process, highly ionized heavy ions in the solar wind collide with neutral atoms or molecules, capturing one or more electrons. The captured electron initially occupies an excited state, and emits soft X-ray or ultraviolet photons as it cascades down to the ground state. Since its initial discovery in observations of Comet Hyakutake in 1996 (\cite{Cravens97}), SWCX emission has been detected by various X-ray observatories, including Chandra, XMM-Newton, and Suzaku (e.g., \cite{Snowden+04}; \cite{Wargelin+04}; \cite{Fujimoto+07}; \cite{Ezoe+10}; \cite{Ezoe+11}; \cite{Ishikawa+13}; \cite{Ishi+19}; \cite{Asakura+21}). Diffuse SWCX emission originating from the solar system is widely recognized as a major component of the soft X-ray background in astronomical observations. However, this emission can also serve as a remote diagnostic of the solar wind plasma, as the energies and intensities of the emission lines depend directly on the ion species and their charge states.

The Xtend instrument on board the X-ray Imaging and Spectroscopy Mission (XRISM: \cite{Tashiro+25}), launched in 2023, possesses capabilities well suited for observing SWCX emission. Xtend consists of four X-ray CCD cameras, providing a wide field of view of $\sim 38.5' \times 38.5'$ and a low instrumental background (\cite{Noda+25_xtend}; \cite{Uchida+25}). These features enable Xtend to detect diffuse X-ray emission from the geocorona and the magnetosphere with high sensitivity. Since XRISM commenced operations near the maximum of the solar cycle, multiple SWCX events were detected soon after the start of the mission. Solar activity was particularly high in 2024; the Hinode/XRT flare catalog lists 53 X-class flares during the year,\footnote{\sloppy \url{https://xrt.cfa.harvard.edu/flare_catalog/2024.html}} including major events in 2024 May and October (\cite{Hayakawa+25}; \cite{Matsumoto+25}). The SWCX events associated with these major solar activities are being analyzed in separate studies (Ishi et al.\ in preparation; Kobayashi et al.\ in preparation). In this paper, we focus on the SWCX event observed with XRISM/Xtend in August 2024. We selected this event because, among the SWCX events identified with XRISM/Xtend in 2024, it yielded the second highest photon statistics after the May 2024 event and showed clear emission lines from heavy elements such as Mg and Si. These features make this dataset suitable for deriving ionization temperatures and abundances for individual elements. The event occurred during intense solar activity, including an X1.3-class flare and a large geomagnetic storm with a minimum Disturbance Storm Time (Dst) index of $\simeq -200$~nT. As discussed in detail in Section~\ref{sec:5}, successive arrivals of multiple CMEs during this period also made this dataset suitable for investigating temporal spectral variations associated with distinct CME components. By analyzing the temporal and spectral signatures of the SWCX emission, we discuss the ion composition and ionization state of the solar wind plasma.

\section{Observation and data reduction}\label{sec:2}

\begin{figure}[t]
 \centering
 \includegraphics[width=8cm]{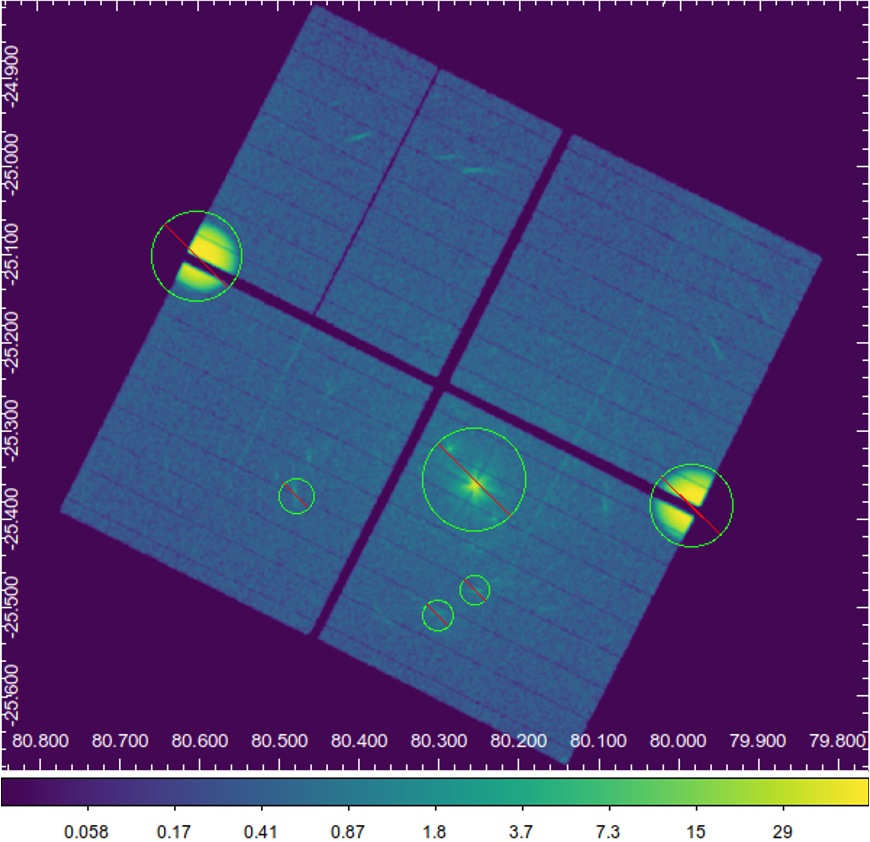}
 \caption{
  Xtend X-ray image in the 0.3--10~keV band. The image was smoothed with a Gaussian kernel of $\sigma=3$ pixels, corresponding to $\sim$5.2$''$. The color scale represents the photon counts per pixel. The axes represent celestial coordinates (J2000.0). The region enclosed in green was excluded from the analysis. Alt text: An X-ray image of the target object.
 }
 \label{fig:image}
\end{figure}

The data used in this study were obtained from an observation of the ultraluminous infrared galaxy IRAS~05189$-$2524 (\cite{Noda+25_iras}), with Obs.~ID 300020010. We conducted the observation using the X-ray CCD camera, Xtend, on board XRISM from 2024 August 9 to August 13 (UT), obtaining an initial effective exposure time of 145.4~ks. The pointing coordinates were $(\rm{RA}, \rm{Dec})_{\mathrm{J2000.0}} = (80.255831^{\circ}, -25.362589^{\circ})$. We calculated the average line-of-sight direction during this observation in Geocentric Solar Ecliptic (GSE) coordinates. The normalized pointing vector in GSE coordinates was $(0.304, -0.591, -0.747)$. This indicates that the line of sight was directed toward the sunward southern high-latitude side of the Earth, near the southern magnetospheric cusp.

A large X1.3-class solar flare occurred on 2024 August 8, approximately one day prior to the observation period. High-resolution observations by Daniel K. Inouye Solar Telescope reported unprecedented fine structures in the flare loops associated with this event (\cite{Tamburri+25}). Following such intense energy release processes, a CME was launched, likely causing significant ionospheric disturbances in addition to a geomagnetic storm in the near-Earth environment. Indeed, an analysis based on Global Navigation Satellite System data reported the occurrence of Large Scale Traveling Ionospheric Disturbances during this period (\cite{Guerra+25}). Consequently, this Xtend observation coincided with significant variations in the near-Earth environment associated with the CME arrival, suggesting a high probability of detecting enhanced SWCX emission from the geocorona.

We performed data processing using HEASoft version 6.35.1, following the standard procedures described in The XRISM Data Reduction Guide (XRISM ABC Guide)\footnote{\sloppy\url{https://heasarc.gsfc.nasa.gov/docs/xrism/analysis/abc_guide/xrism_abc.html}}. First, we applied the \texttt{searchflickpix} command to remove flickering pixels from the Xtend data. Subsequently, we masked unrelated point sources selected with reference to the existing point source catalog\footnote{\sloppy \url{https://simbad.u-strasbg.fr/simbad/sim-fbasic}}. Figure~\ref{fig:image} shows the image of the entire field of view after these initial screenings. Several faint point sources remain visible in the field of view, including both cataloged and uncataloged sources. We confirmed that their contribution to the overall analysis is negligible by verifying that a more stringent point-source mask yields spectral parameters consistent within the statistical uncertainties.

We generated the response files for spectral analysis using the \texttt{xtdrmf} and \texttt{xaarfgen} commands. For the Auxiliary Response File, we assumed uniform emission from a circular region with a radius of $40'$, covering the full field of view. During the spectral fitting, we normalized the flux by the solid angle of the analysis region to obtain the surface brightness of the diffuse emission. We binned the spectra to ensure a minimum of 20 counts per bin.

We generated a light curve in the 3.0--10.0~keV band with a time bin size of 256~s to identify spike-like background increases during the latter part of the observation, likely caused by high-energy particles. We excluded the period from 2024-08-12T07:40:41 to 2024-08-12T10:27:21, which corresponds to approximately two orbital periods around the event time. Furthermore, we removed contamination from solar X-rays scattered by the Earth's atmosphere (Day Earth) and auroral X-rays (Night Earth) with reference to the user manual\footnote{\sloppy\url{https://xrism.isas.jaxa.jp/research/analysis/dataanalyses/ScreeningContami_Xtend_v241216.pdf}}. To eliminate Day Earth contamination, we extracted time intervals where the elevation angle from the bright Earth limb (\texttt{DYE\_ELV}) was $> 60^{\circ}$ throughout the entire observation period. Regarding Night Earth contamination, we confirmed an enhancement likely due to aurora after 06:00 UT on 2024 August 12. Therefore, specifically for the data after this time, we applied an additional filter to select time intervals where the elevation angle from the dark Earth limb (\texttt{NTE\_ELV}) was $> 15^{\circ}$. After applying these filtering criteria, the net exposure time was 71.8~ks, corresponding to 49.4\% of the initial exposure. Unless otherwise noted, all error bars in the figures and uncertainties in the parameters presented in this paper correspond to the $1\sigma$ confidence level.

\section{Light curve}\label{sec:3}

\begin{figure*}[t]
 \centering
 \includegraphics[width=0.8\textwidth]{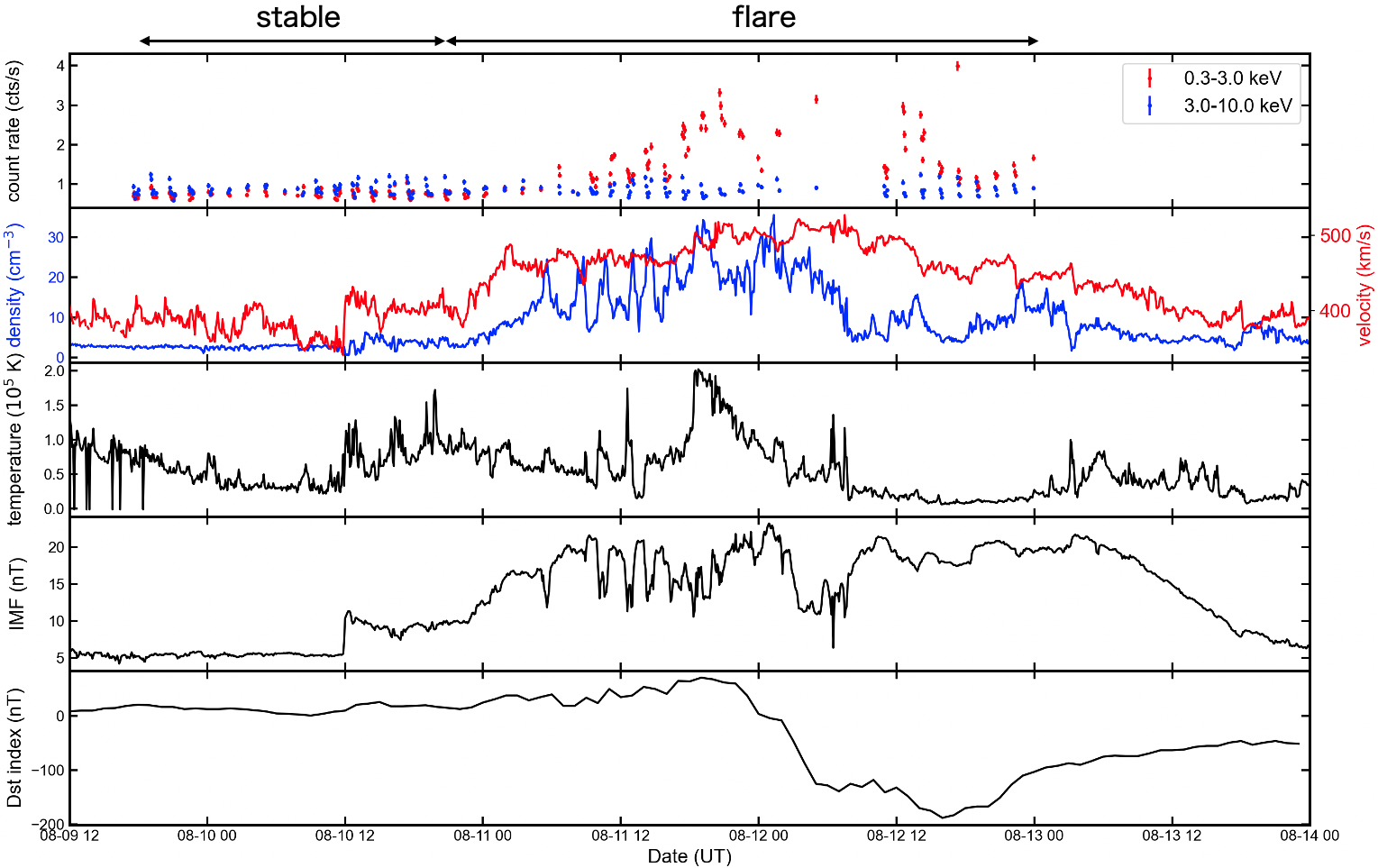}
 \caption{
  Temporal variations of dataset parameters. From top to bottom: Xtend light curves in the 0.3--3.0~keV (red) and 3.0--10.0~keV (blue) bands; solar wind proton density (blue) and velocity (red); proton temperature; IMF magnitude; and the Dst index. Solar wind data were obtained from ACE/SWEPAM. The Dst index is provided by the WDC for Geomagnetism, Kyoto, Japan. Alt text: Five stacked line graphs sharing a horizontal time axis.
 }
 \label{fig:lightcurve}
\end{figure*}

Figure \ref{fig:lightcurve} shows the light curves obtained with Xtend in the soft X-ray (0.3--3.0~keV) and hard X-ray (3.0--10.0~keV) bands. The data were binned with a time bin size of 256~s. In the latter part of the observation, a significant enhancement was observed in the soft X-ray band. Accordingly, based on this soft X-ray variability, we defined the period prior to the enhancement (from 17:34:01 UT on 2024 August 9 to 21:20:41 UT on August 10) as the ``stable'' period and the period covering the enhancement (from 21:20:41 UT on August 10 to 00:38:59 UT on August 13) as the ``flare'' period. We estimated the clear onset of the soft X-ray enhancement as the first time when the 0.3--3.0~keV count rate exceeded the mean count rate before the enhancement by more than $1\sigma$ for four consecutive 256~s bins. We conservatively set the boundary between the ``stable'' and ``flare'' periods to an earlier GTI boundary in order to exclude the possible early phase of the enhancement from the ``stable'' period. Applying the same criterion to the 3.0--10.0~keV light curve did not identify a corresponding enhancement.

In the soft X-ray band, the average count rates were $0.747 \pm 0.054~\mathrm{cts~s^{-1}}$ during the ``stable'' period and $1.605 \pm 0.080~\mathrm{cts~s^{-1}}$ during the ``flare'' period, confirming an increase by a factor of more than two in the ``flare'' period compared to the ``stable'' period. On the other hand, in the hard X-ray band, the average count rates were $0.864 \pm 0.058~\mathrm{cts~s^{-1}}$ and $0.858 \pm 0.058~\mathrm{cts~s^{-1}}$ for the ``stable'' and ``flare'' periods, respectively, showing no significant variation.

Figure \ref{fig:lightcurve} also displays the temporal variations of solar wind density, velocity, temperature, and interplanetary magnetic field (IMF) magnitude obtained by the Solar Wind Electron Proton Alpha Monitor (SWEPAM) on board the ACE satellite\footnote{\sloppy \url{https://izw1.caltech.edu/ACE/ASC/level2/lvl2DATA_SWEPAM.html}}. Around 12:00:00 UT on 2024 August 10, discontinuous changes were observed in the velocity, temperature, and IMF, suggesting the passage of a shock wave associated with a CME near Earth at this time. Subsequently, these parameters gradually increased, and the IMF exhibited significant fluctuations during the first half of the ``flare'' period. These fluctuations are likely attributable to the highly turbulent sheath region formed behind the shock wave. Furthermore, in the latter half of the ``flare'' period, decreases in density, velocity, and temperature, along with a smoothing of the IMF, were observed. These signatures suggest the passage of the CME ejecta (main body), which is often identified as a magnetic cloud, near Earth.

At the same time, a significant decrease was confirmed in the Dst index provided by the World Data Center for Geomagnetism, Kyoto, Japan\footnote{\url{https://wdc.kugi.kyoto-u.ac.jp/dstdir/}}. The Dst index is an indicator of geomagnetic disturbance derived from the average of the horizontal component at low- and mid-latitude geomagnetic observatories. A large negative excursion in the Dst index indicates the occurrence of a geomagnetic storm, signifying the arrival of the CME body at the Earth's magnetosphere. These results confirm a typical CME arrival process, suggesting that the soft X-ray enhancement observed by Xtend during the ``flare'' period is likely SWCX emission originating from the CME.

\begin{figure*}[t]
 \centering
 \includegraphics[width=0.8\textwidth]{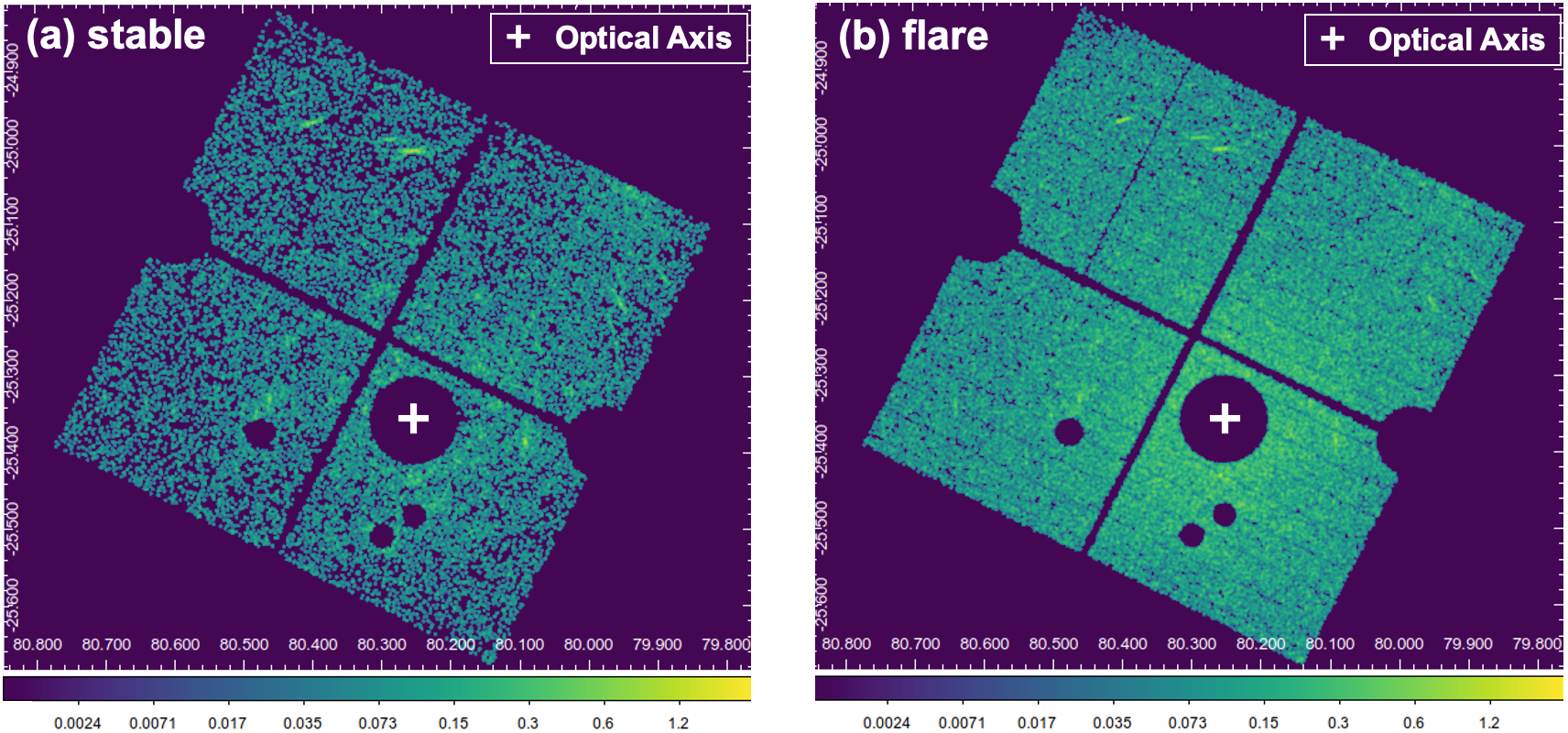}
 \caption{
  Xtend images in the 0.3--3.0~keV band obtained during (a) the ``stable'' period and (b) the ``flare'' period. The color bar represents the count rate in units of $10^{-5}$~counts~s$^{-1}$~pixel$^{-1}$. The axes represent celestial coordinates, and the cross mark indicates the position of the optical axis. The images were smoothed with a Gaussian kernel of $\sigma=3$ pixels. A global increase in the count rate is evident during the ``flare'' period. Alt text: Two X-ray images.
 }
 \label{fig:image_comparison}
\end{figure*}

Figure \ref{fig:image_comparison} shows the Xtend images in the 0.3--3.0~keV band for the ``stable'' and ``flare'' periods. The count rate in the ``flare'' period is significantly higher than that in the ``stable'' period. The image during the ``flare'' period shows an enhancement over a broad region of the field of view, rather than a localized structure. Its radial profile is consistent with the vignetting expected for diffuse X-ray emission entering through the telescope optics, supporting the interpretation that the enhancement is not dominated by point sources or detector background. A quantitative comparison between the radial profile of the enhancement and the expected vignetting profile is presented in Appendix. We further investigate the origin of this emission through detailed spectral analysis in the following section.

\section{Spectrum}\label{sec:4}
\begin{figure}[t]
 \centering
 \includegraphics[width=8cm]{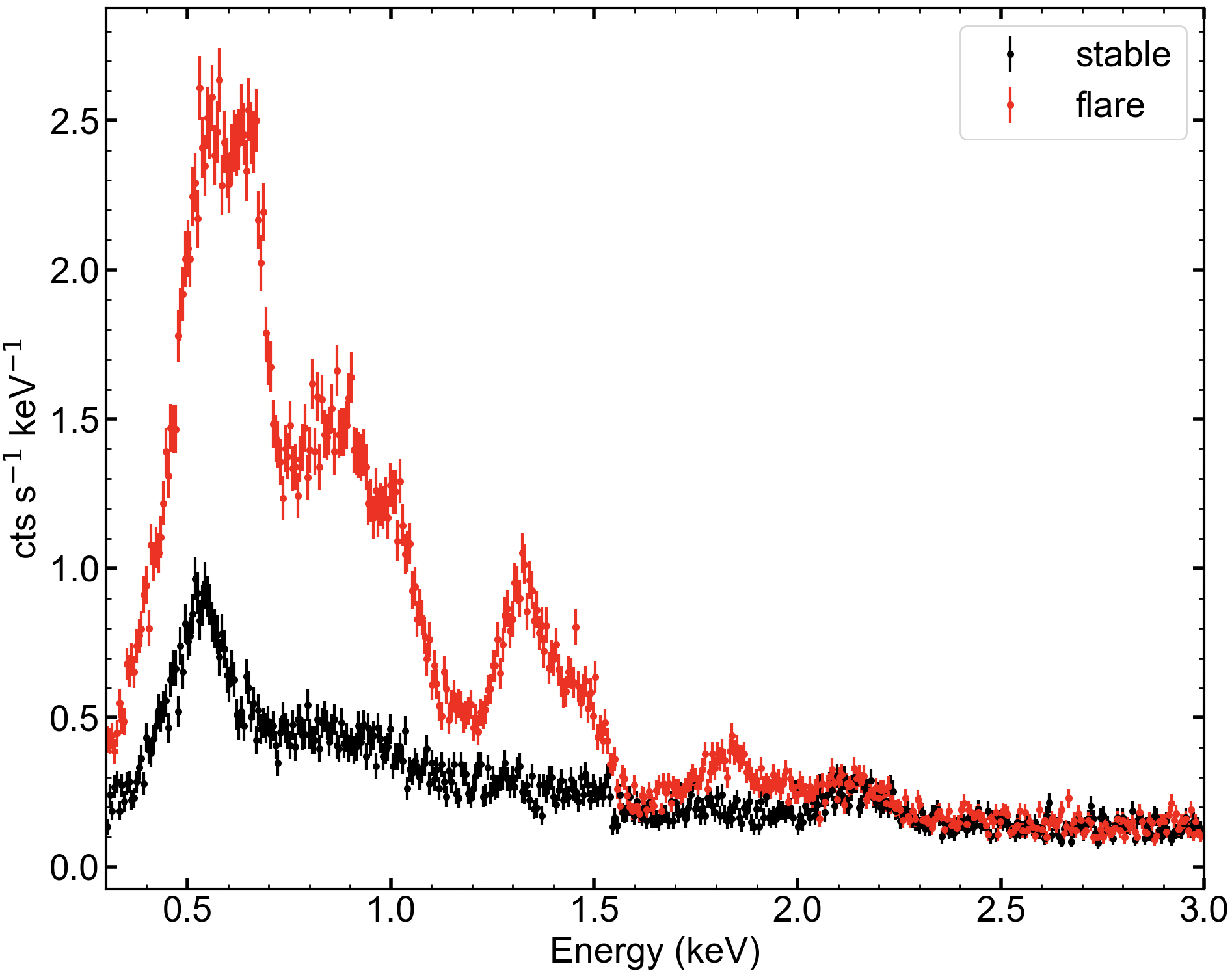}
 \caption{Xtend spectra obtained during the ``stable'' (black) and ``flare'' (red) periods. Alt text: A line graph displaying two X-ray spectra.
 }
 \label{fig:sp1}
\end{figure}

Figure \ref{fig:sp1} shows the spectra in the 0.3--3.0~keV band obtained during the ``stable'' and ``flare'' periods. The spectrum during the ``stable'' period consists of sky background components (Galactic and extragalactic diffuse emission), the non-X-ray background (NXB), and the fluorescent O K$\alpha$ line produced by solar X-rays scattered by neutral oxygen in the Earth's atmosphere. In contrast to the ``stable'' spectrum, the ``flare'' spectrum exhibits multiple emission lines likely associated with SWCX. 

To assess whether the ``stable'' spectrum was suitable as the reference spectrum for background subtraction, we examined the temporal stability of its constituent components. The sky background components are considered to exhibit little temporal variation and remain substantially constant throughout the observation (\cite{Ezoe+11}). We compared the NXB levels between the ``stable'' and ``flare'' periods by fitting their 7.0--15.0~keV spectra with the average Xtend NXB model,\footnote{\sloppy \url{https://heasarc.gsfc.nasa.gov/docs/xrism/analysis/nxb/xtend_nxb_spectral_model.html}} allowing only its normalization to vary. The model reproduced the spectral shape in both periods. After correcting for the areas of the extraction regions, the normalization factors were \(1.1721 \pm 0.0073\) and \(1.1707 \pm 0.0070\) for the ``stable'' and ``flare'' periods, respectively, and were consistent within the statistical uncertainties. We therefore adopted the ``stable'' spectrum as the reference for background subtraction in all spectral analyses presented in this paper. As an additional check, we shifted the boundary between the ``stable'' and ``flare'' periods by approximately $\pm 5$~hr and confirmed that the resulting spectral parameters remained consistent within the statistical uncertainties.

\begin{figure}[t]
 \centering
 \includegraphics[width=8cm]{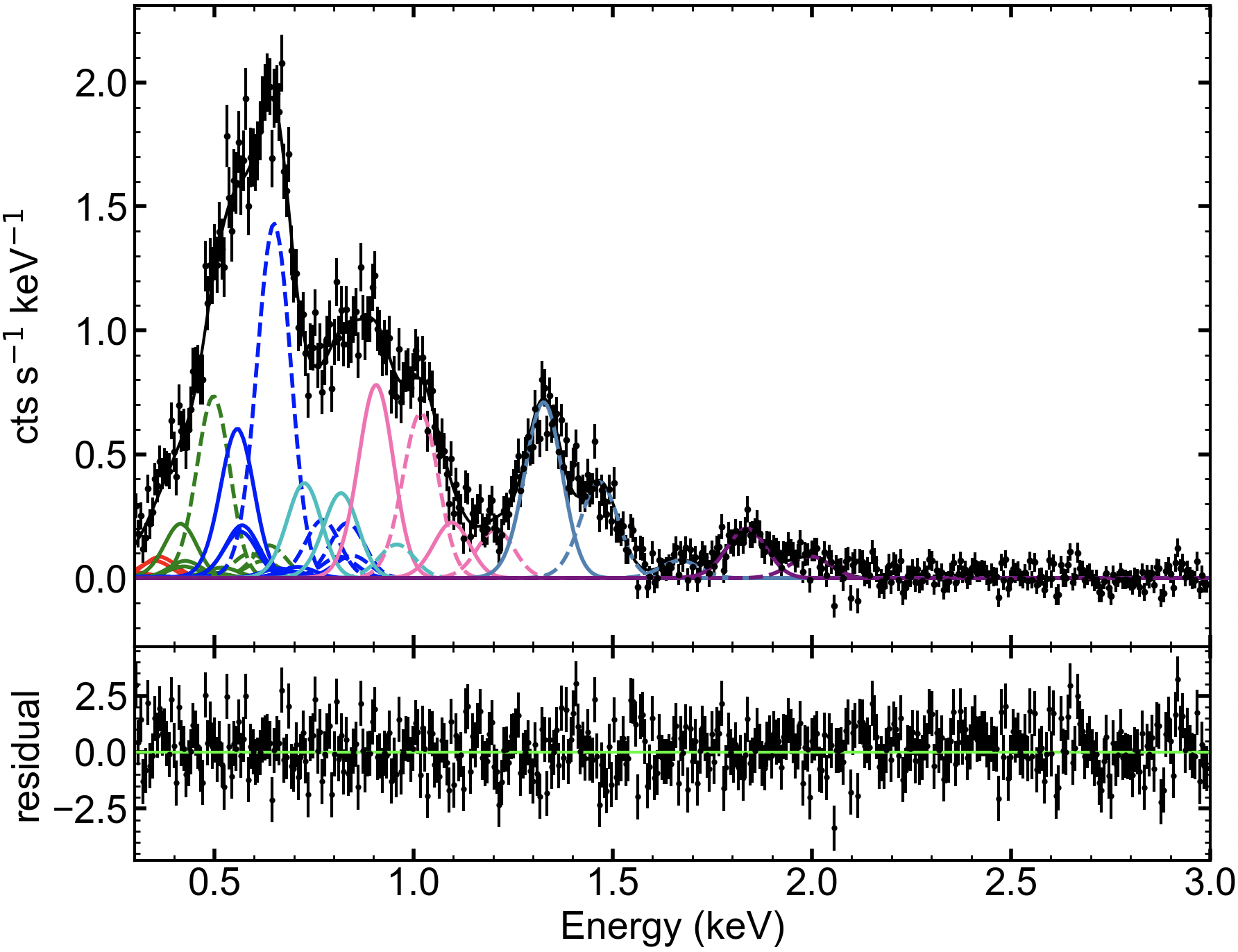}
 \caption{
  Xtend spectrum during the ``flare'' period fitted with Gaussian models. The spectrum of the ``stable'' period was subtracted as background. The parameters of the model are listed in Table \ref{tab:gauss_fit}. The colors of the emission lines represent C (red), N (green), O (blue), Fe (cyan), Ne (pink), Mg (slate blue), and Si (purple). The linestyles indicate He-like ions (except for Fe) and Fe$_{\rm XVII}$ (dotted lines), and H-like ions (except for Fe) and Fe$_{\rm XX}$ (solid lines). Alt text: A line graph showing the X-ray spectrum and the best fit of multiple Gaussian models, along with a residual panel.
 }
 \label{fig:sp2}
\end{figure}

We fitted the background-subtracted spectrum of the ``flare'' period with Gaussian models using XSPEC. The resulting spectrum and the best-fit model are shown in Figure \ref{fig:sp2}. The line widths were fixed at 0~eV, and the line center energies were fixed to theoretical values. For C, N, and O, each ionization state was represented by a set of Gaussian lines, and the relative intensities of the transitions belonging to the same ion were fixed using the ratios given in Table~2 of \citet{Bodewits+07} for collision velocities of 200, 400, 600, 800, and 1000~$\mathrm{km\,s^{-1}}$. Since the average solar wind speed during the ``flare'' period was 472~$\mathrm{km\,s^{-1}}$, we adopted the ratios for 400~$\mathrm{km\,s^{-1}}$. For Fe, Ne, Mg, and Si, we constructed Gaussian line sets referring to past detections (\cite{Carter+10}; \cite{Ishi+19}; \cite{Asakura+21}).

Table \ref{tab:gauss_fit} lists the best-fit parameters obtained from the Gaussian fitting. In addition to C, N, and O lines often detected in previous observations, emission lines in the 1--2~keV band, such as Mg$_{\rm XI}$ (1.33~keV) and Si$_{\rm XIII}$ (1.84~keV), which typically suffer from limited statistics, were detected. Based on the fit improvement ($\Delta\chi^2$) for one additional degree of freedom, they achieved remarkably high significances of $\sim 35\sigma$ ($\Delta\chi^2 \approx 1266$) and $\sim 15\sigma$ ($\Delta\chi^2 \approx 230$), respectively. The grasp, defined as the effective area integrated over the field of view, of Xtend in the 1--2~keV band is approximately 2--3 times that of the Suzaku satellite. This high sensitivity likely enabled the detection of emission lines in the 1--2~keV band with high statistical quality. On the other hand, the S$_{\rm XVI}$ line at 2.62~keV, which was detected in the SWCX event reported by \citet{Asakura+21}, was not significantly detected in our observation. This is likely due to the difference in the solar flare scale. The event analyzed by \citet{Asakura+21} was associated with an X6-class flare, which is significantly more intense than the X1.3-class flare in the present study.

\begin{table}[t]
 \centering
 \caption{\raggedright Best-fit parameters of the Gaussian models shown in Figure~\ref{fig:sp2}. For C, N, and O, only the strongest line of each ionization state is listed here. \par}
 \begin{tabular}{lccc}
  \hline
  Ion & $E_\mathrm{c}$ & Normalization & $f_\mathrm{X}$ \\
  \hline
  C$_{\rm V}$ & 299 & $191.3 \pm 26.7$ & $3.9 \times 10^{-11}$ \\
  C$_{\rm VI}$ & 367 & 0.6 (<2.8)  & $1.4 \times 10^{-13}$ \\
  N$_{\rm VI}$ & 420 & $13.3 \pm 1.4$  & $3.8 \times 10^{-12}$ \\
  N$_{\rm VII}$ & 500 & $18.6 \pm 1.1$  & $6.4 \times 10^{-12}$ \\
  O$_{\rm VII}$ & 561 & $15.2 \pm 0.8$  & $5.8 \times 10^{-12}$ \\
  O$_{\rm VIII}$ & 653 & $22.2 \pm 0.7$  & $9.9 \times 10^{-12}$ \\
  Fe$_{\rm XVII}$ & 730 & $4.7 \pm 0.5$  & $2.3 \times 10^{-12}$ \\
  Fe$_{\rm XVII}$ & 820 & $3.5 \pm 0.4$  & $2.0 \times 10^{-12}$ \\
  Ne$_{\rm IX}$ & 920 & $7.1 \pm 0.4$  & $4.4 \times 10^{-12}$ \\
  Fe$_{\rm XX}$ & 960 & $1.2 \pm 0.3$  & $7.7 \times 10^{-13}$ \\
  Ne$_{\rm X}$ & 1022 & $5.5 \pm 0.3$  & $3.8 \times 10^{-12}$ \\
  Ne$_{\rm IX}$ & 1100 & $1.8 \pm 0.2$  & $1.3 \times 10^{-12}$ \\
  Ne$_{\rm X}$ & 1220 & $1.5 \pm 0.2$  & $1.2 \times 10^{-12}$ \\
  Mg$_{\rm XI}$ & 1330 & $5.1 \pm 0.2$  & $4.7 \times 10^{-12}$ \\
  Mg$_{\rm XII}$ & 1470 & $2.8 \pm 0.2$  & $2.8 \times 10^{-12}$ \\
  Mg$_{\rm XI}$ & 1690 & $0.7 \pm 0.2$  & $7.9 \times 10^{-13}$ \\
  Si$_{\rm XIII}$ & 1840 & $2.0 \pm 0.2$  & $2.5 \times 10^{-12}$ \\
  Si$_{\rm XIV}$ & 2000 & $0.9 \pm 0.2$  & $1.2 \times 10^{-12}$ \\
  \hline
  $\chi^2/\mathrm{d.o.f.}$ & 507.85/431 \\
   
  \hline
 \end{tabular}
 \label{tab:gauss_fit}
 \footnotesize
 \\$E_\mathrm{c}$ represents the line center energy in units of eV. Normalization represents the line intensity in units of $\mathrm{ph\,s^{-1}\,cm^{-2}\,sr^{-1}}$ (Line Units; LU). $f_\mathrm{X}$ represents the X-ray energy flux in units of $\mathrm{erg\,s^{-1}\,cm^{-2}}$. All the line widths were fixed at 0~eV.
\end{table}

\section{Multiple CME events}\label{sec:5}
\subsection{Separation of events}\label{ssec:51}
\begin{figure*}[t]
 \centering
 \includegraphics[width=0.8\textwidth]{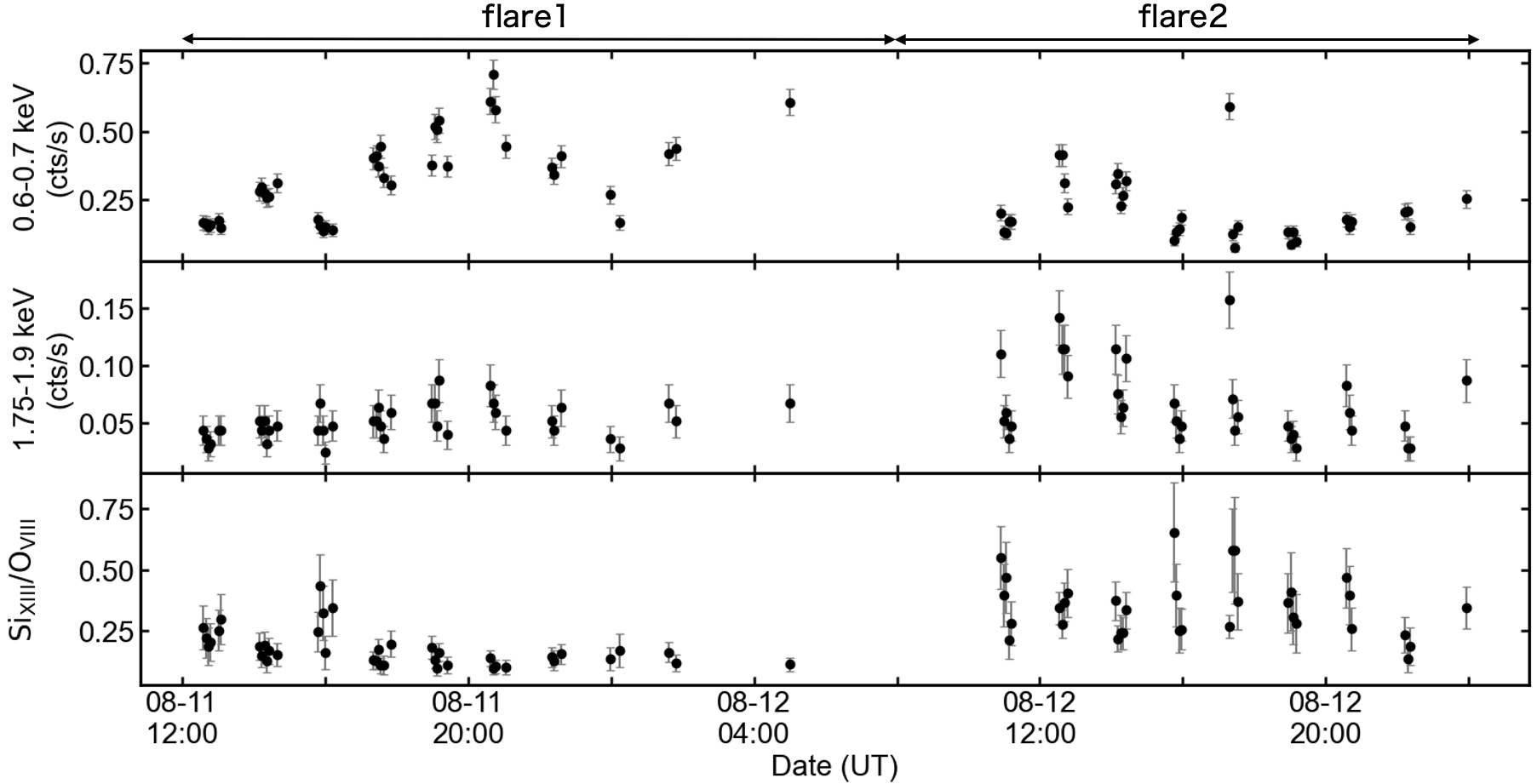}
 \caption{
  Xtend light curves in the 0.6--0.7~keV (O$_{\rm VIII}$) and 1.75--1.9~keV (Si$_{\rm XIII}$) bands, and their ratio. The ``flare1'' and ``flare2'' periods were defined based on the CME arrival times. Alt text: Three line graphs showing X-ray light curves and their ratio, extracted for the flare 1 and flare 2 periods.
 }
 \label{fig:line_ratio}
\end{figure*}
According to the Geostationary Operational Environmental Satellite (GOES)\footnote{\sloppy \url{https://www.swpc.noaa.gov/products/goes-x-ray-flux}}, multiple M-class and X-class flares occurred prior to the start of the Xtend observation. Specifically, for the CMEs associated with the M1.9, M4.0, and M5.0 class flares that occurred at 03:24, 14:24, and 19:36 UT on 2024 August 7, respectively, NASA Community Coordinated Modeling Center (CCMC)\footnote{\url{https://kauai.ccmc.gsfc.nasa.gov/CMEscoreboard/PreviousPredictions/2024}} estimated that their shock waves would arrive in the near-Earth environment around 11:55 UT on 2024 August 10.

As shown in Figure \ref{fig:lightcurve}, the soft X-ray enhancement began around 12:00 UT on 2024 August 11, approximately one day after the predicted shock arrival. Since this timing is consistent with the arrival of these CMEs, it is highly likely that they contributed to the SWCX enhancement. Meanwhile, regarding the CME associated with the X1.3-class flare on 2024 August 8 (19:36 UT), the shock wave and the CME main body were predicted by the CCMC to arrive in the near-Earth environment at 12:00 UT on August 11 and 07:44 UT on August 12, respectively. These predictions suggest that during the period from August 11 to August 12, SWCX enhancements caused by at least two distinct CME events likely occurred successively and superimposed on each other. Therefore, based on the arrival timing of the main body of the latter CME, we defined the period from 12:00 UT on August 11 to 06:00 UT on August 12 as ``flare1'' (enhancement by the first group of CMEs) and the period from 06:00 UT on August 12 to 00:00 UT on August 13 as ``flare2'' (enhancement by the latter CME), and performed analysis for each period.

Figure \ref{fig:line_ratio} shows the light curves in the 0.6--0.7~keV (O$_{\rm VIII}$) and 1.75--1.9~keV (Si$_{\rm XIII}$) bands, along with the temporal variation of their intensity ratio (line ratio). The average count rate of O$_{\rm VIII}$ was $0.264 \pm 0.005$ in the ``flare1'' period and $0.167 \pm 0.005$ in the ``flare2'' period, showing a decrease of approximately 33\% from ``flare1'' to ``flare2''. On the other hand, the average count rate of Si$_{\rm XIII}$ was $0.046 \pm 0.002$ in the ``flare1'' period and $0.054 \pm 0.003$ in the ``flare2'' period, confirming an increase of approximately 17\%. We fitted the line ratio with a constant model over the entire ``flare1''+``flare2'' interval. The result yielded $\chi^2/\mathrm{d.o.f.} = 168.80/72$ and a p-value of $9.55 \times 10^{-10}$, rejecting the constant model with a confidence level of $>99\%$. In contrast, we fitted the line ratio separately with constant models for ``flare1'' and ``flare2'', and obtained $\chi^2/\mathrm{d.o.f.} = 39.52/39$ ($p = 0.447$) and $39.62/32$ ($p = 0.167$), respectively, showing that each period is individually consistent with a constant value within the statistical uncertainties. These results suggest that the ion compositions of the CMEs contributing to the enhancements in the ``flare1'' and ``flare2'' periods were different. Furthermore, since the ionization potential of Si is significantly higher than that of O, the increase in the Si$_{\rm XIII}$ count rate implies that the ions contained in the latter CME had a higher ionization temperature than those in the former CMEs. To quantitatively evaluate the difference in the ionization temperatures and physical properties suggested by the line ratios, we conducted spectral analysis using a spectral model for charge exchange emission.

\subsection{Spectral modeling}\label{ssec:52}
\begin{figure}[t]
 \centering
 \includegraphics[width=8cm]{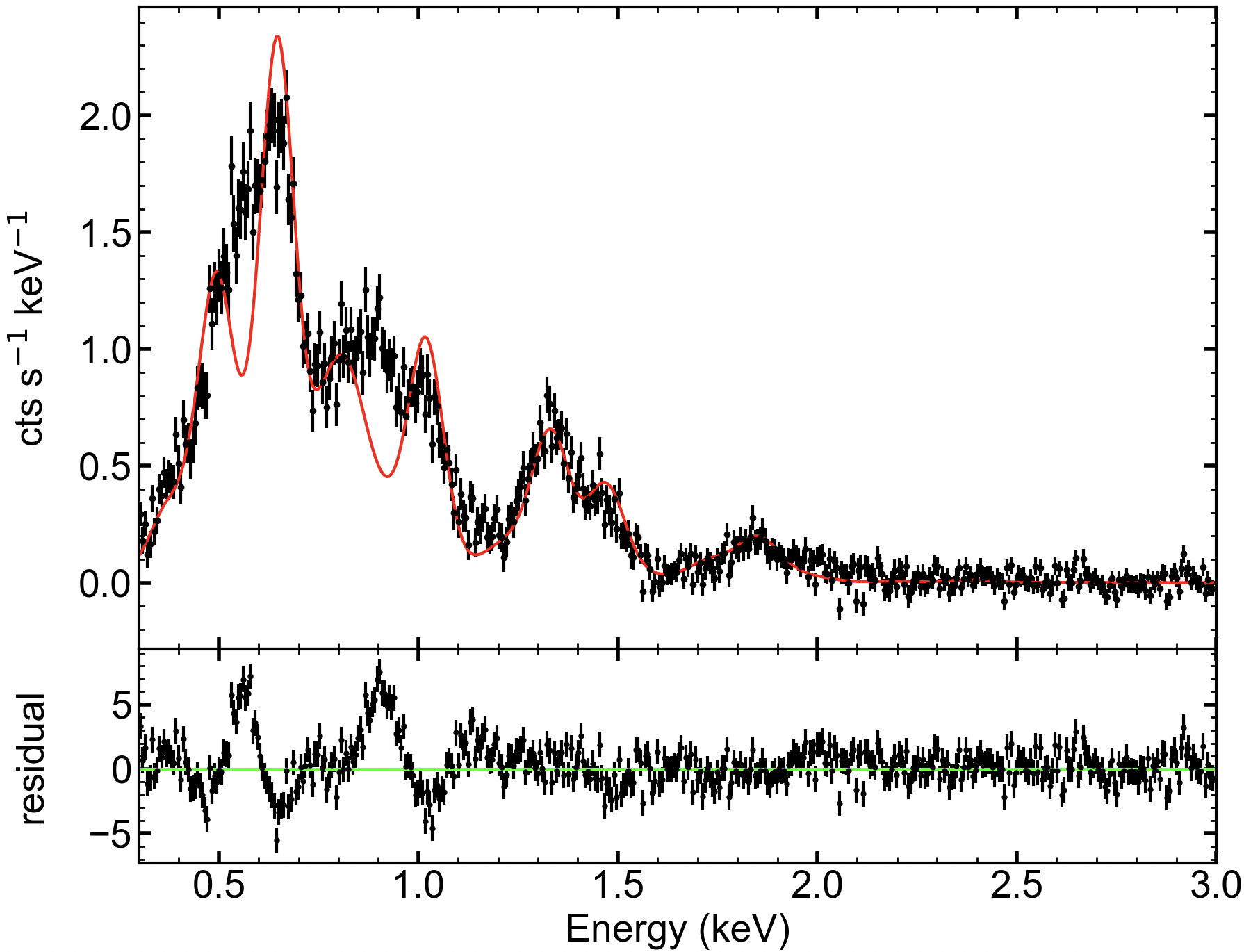}
 \caption{
  Comparison of the spectrum for the entire ``flare'' period (black) with the \texttt{vacx2} single-temperature model (red) (top), and the residuals (bottom). As indicated by the large residuals, the single-temperature model fails to reproduce the observed spectrum. Alt text: A line graph comparing an X-ray spectrum with a model fit, along with a residual panel.
 }
 \label{fig:sp3}
\end{figure}

\begin{figure*}[t]
 \centering
 \includegraphics[width=0.8\textwidth]{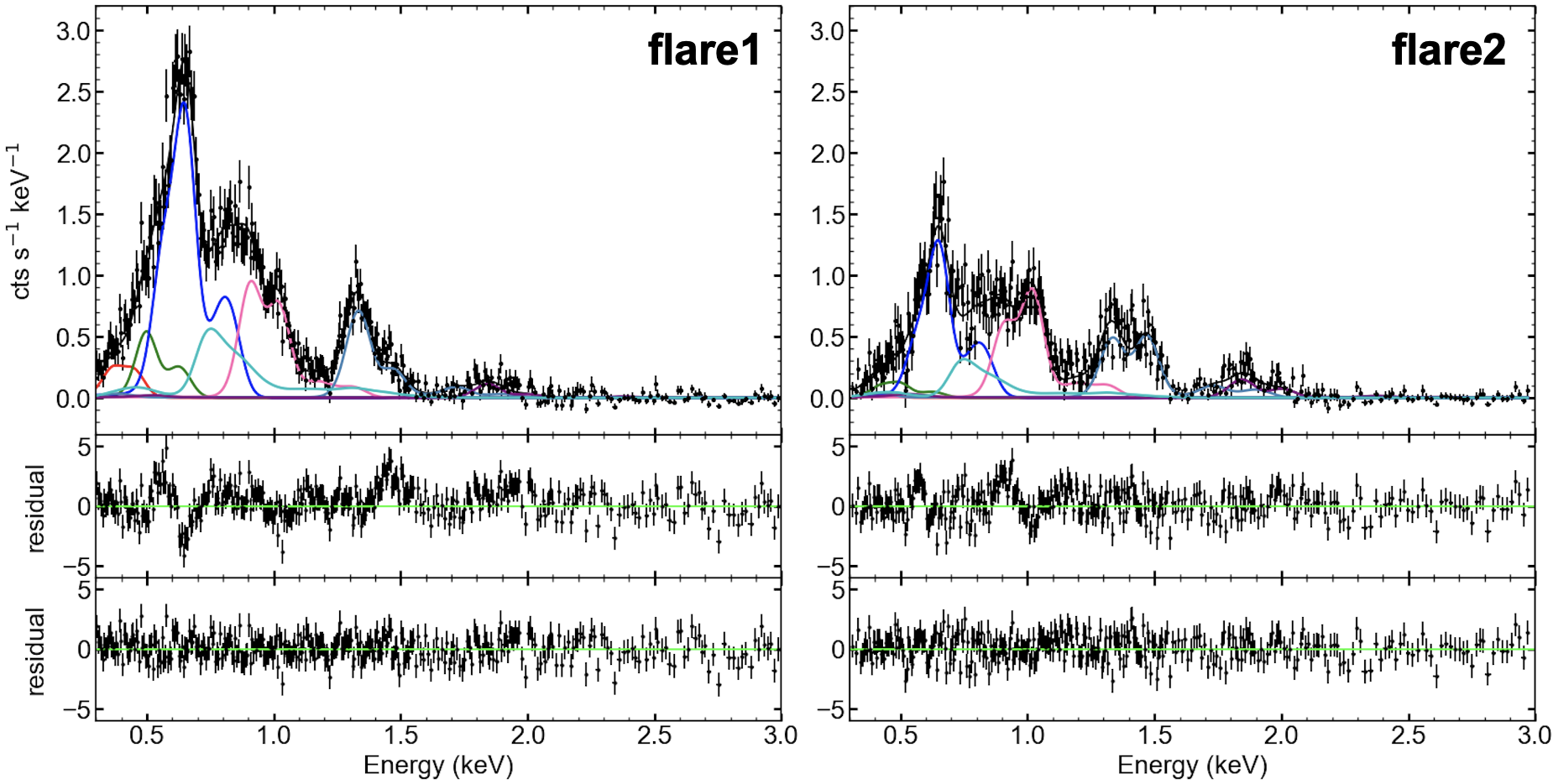}
 \caption{
  Spectra of the ``flare1'' period (left) and ``flare2'' period (right). The top panels show the data fitted with the \texttt{vacx2} multi-temperature model, where the colors of the components represent C (red), N (green), O (blue), Fe (cyan), Ne (pink), Mg (steel blue), and Si (purple). The middle and bottom panels display the residuals from the \texttt{vacx2} single-temperature model and the multi-temperature model, respectively. The multi-temperature model yields significantly smaller residuals compared to the single-temperature model. Alt text: Two sets of X-ray spectra and model fits, each accompanied by two residual panels.
 }
 \label{fig:sp4}
\end{figure*}

We performed spectral analysis using the AtomDB Charge Exchange model version 2.3.1 (ACX2 model: \cite{Mullen+16}; \cite{Mullen+17}; \cite{Cumbee+17}). The ACX2 model includes the relative abundance to the solar photosphere, plasma ionization temperature, collision velocity between plasma and neutrals, and normalization as its main parameters. In this analysis, we applied the \texttt{vacx2} model, a variant of the ACX model that allows abundances to be set independently for each element. First, we applied the \texttt{vacx2} model with a single temperature to the spectrum of the entire ``flare'' period. In this model, the abundances of C, N, O, Ne, Mg, Si, and Fe, a common ionization temperature, and the normalization were treated as free parameters. The O abundance was fixed at 1, and the collision velocity was fixed at 500~$\mathrm{km\,s^{-1}}$, based on the average solar wind proton velocity obtained by the ACE satellite during this period. As shown in Figure \ref{fig:sp3}, large residuals were observed in the 0.3--1.2~keV band, and the obtained statistic was $\chi^2/\mathrm{d.o.f.} = 1507.64/441$. A similar trend was reported by \citet{Zhou+23}, who obtained a reduced chi-squared value of $> 3$ when applying a single-temperature ACX model to Suzaku observations. Thanks to the high photon statistics of Xtend, our observation clearly confirmed this discrepancy. Furthermore, a recent Xtend observation of another SWCX event in May 2024 also found the single-temperature model to be insufficient (Kobayashi et al.\ in preparation). These results confirm that the single-temperature model cannot sufficiently reproduce the observed spectra, suggesting that the ionization temperatures of light elements (C, N, O) and heavy elements (Ne, Mg, Si, Fe) differ significantly.

Considering previous reports from solar wind observations by the ACE and Ulysses satellites that ionization states differ for each element (e.g., \cite{Geiss+95}; \cite{von Steiger+11}), we constructed a multi-temperature model in which the temperature is set independently for each element, following the method of \citet{Asakura+21}. Specifically, we used seven \texttt{vacx2} components, one for each element, setting the abundance of non-designated elements to 0 in each component. The ionization temperature parameter for each model was allowed to vary freely and independently, thereby reproducing different temperatures for each element. The O abundance was fixed at 1 as before, while the abundances of other elements were allowed to vary. Furthermore, only the normalization of the O component was allowed to vary freely, while the normalizations of the other elements were linked to the O value.

Figure \ref{fig:sp4} shows the fitting results of applying the multi-temperature model to the ``flare1'' and ``flare2'' periods. The residuals are clearly reduced compared to the single-temperature model, and the obtained statistics were $\chi^2/\mathrm{d.o.f.} = 336.57/323$ for ``flare1'' and $344.61/300$ for ``flare2''. The significant improvement in spectral reproducibility demonstrates the validity of the multi-temperature model. Table \ref{tab:acx_both_flares} summarizes the main parameters obtained from the \texttt{vacx2} multi-temperature model fitting. The C and N parameters are less well constrained than those of the other elements, likely because their lines are not well separated in energy in the lower part of the spectrum at the CCD resolution. We discuss the physical implications of these parameters in the following section.

\section{Discussion}\label{sec:6}
\begin{table*}[t]
 \centering
 \caption{Best-fit parameters of the \texttt{vacx2} multi-temperature model for the ``flare1'' and ``flare2'' period spectra. Abundances are relative to the solar photospheric values and are normalized to oxygen. The normalization parameter was linked across all elemental components within each period.}
 \label{tab:acx_both_flares}
 \begin{tabular}{lcccc}
  \hline
  & \multicolumn{2}{c}{``flare1''} & \multicolumn{2}{c}{``flare2''} \\
  Element & Abundance [$(\mathrm{X}/\mathrm{O})/(\mathrm{X}/\mathrm{O})_\odot$] & $kT$ [keV] & Abundance [$(\mathrm{X}/\mathrm{O})/(\mathrm{X}/\mathrm{O})_\odot$] & $kT$ [keV] \\
  \hline
  C  & $4.16$ $(<10)$    & $0.10$ $(<0.36)$  & $0.26$ $(<10)$     & $<86^\dagger$ \\
  N  & $1.96$ $(<4.57)$  & $<86^\dagger$     & $3.31$ $(<10)$     & $0.14 \pm 0.05$ \\
  O  & 1 (fixed)         & $0.26 \pm 0.01$   & 1 (fixed)          & $0.28 \pm 0.02$ \\
  Ne & $1.70 \pm 0.19$   & $0.39 \pm 0.01$   & $2.52 \pm 0.35$    & $0.45 \pm 0.02$ \\
  Mg & $3.06 \pm 0.26$   & $0.57 \pm 0.02$   & $4.27 \pm 0.47$    & $0.76 \pm 0.02$ \\
  Si & $0.59 \pm 0.20$   & $0.85 \pm 0.12$   & $1.41 \pm 0.29$    & $1.00 \pm 0.08$ \\
  Fe & $8.45 \pm 0.98$   & $0.68 \pm 0.05$   & $<10^\dagger$      & $0.66 \pm 0.08$ \\
  \hline
  Normalization & \multicolumn{2}{c}{$6.86 \pm 0.34$} & \multicolumn{2}{c}{$3.27 \pm 0.31$} \\
  $\chi^2/\mathrm{d.o.f.}$ & \multicolumn{2}{c}{336.57/323} & \multicolumn{2}{c}{344.61/300} \\
  \hline
  \multicolumn{5}{l}{\footnotesize $^\dagger$ Unconstrained upper limit.} \\
 \end{tabular}
\end{table*}

\subsection{Ionization temperature vs flare intensity}\label{ssec:61}
\begin{figure}[t]
 \centering
 \includegraphics[width=8cm]{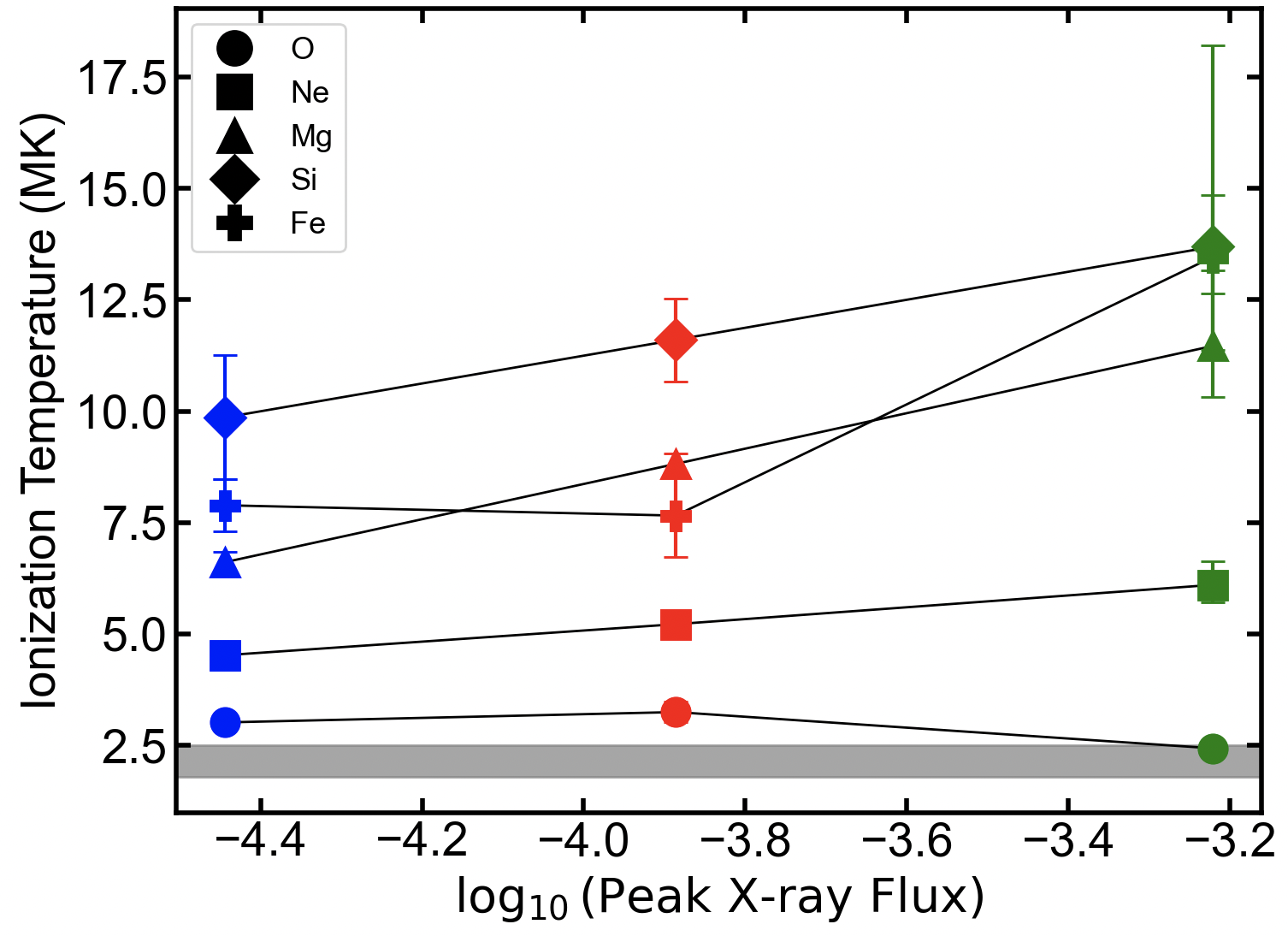}
 \caption{
  Relationship between the peak X-ray flux of the solar flare corresponding to each observation obtained by GOES and the ionization temperature of each element obtained from the \texttt{vacx2} multi-temperature model fitting. Blue represents ``flare1'', red represents ``flare2'', and green represents \citet{Asakura+21}. The gray area indicates the typical range of oxygen ionization temperatures during flares based on ACE satellite observations (\cite{Reinard05}). Alt text: A line graph showing the relationship between the solar flare peak X-ray flux and the ionization temperatures of various elements.
 }
 \label{fig:temp_intensity}
\end{figure}

Excluding C and N, which have large uncertainties, the ionization temperature parameters for the same element were consistently higher in ``flare2'' than in ``flare1''. This suggests that the CME plasma contributing to the ``flare2'' enhancement was ionized in a hotter environment.

We focused on the intensity of the solar flares originating the CMEs as a factor causing the difference in ionization temperatures. Figure \ref{fig:temp_intensity} plots the ionization temperatures of each element (excluding C and N) against the peak X-ray flux of the source flare for ``flare1'', ``flare2'', and the SWCX event associated with an X6-class flare (\cite{Asakura+21}). For ``flare1'', since three CME groups are considered to be mixed, we used the average value of their flare intensities for the horizontal axis. The gray area indicates the typical range of oxygen ionization temperatures estimated from the $\mathrm{O^{+7}/O^{+6}}$ ratio during C- to X-class flare events observed by the ACE satellite (\cite{Reinard05}). Compared with the typical ACE value of 1.8--2.5~MK, the oxygen ionization temperatures obtained for ``flare1'', ``flare2'', and \citet{Asakura+21} are $3.0 \pm 0.1$, $3.2 \pm 0.23$, and $2.4 \pm 0.11$~MK, respectively. The value for \citet{Asakura+21} is within the typical ACE range, whereas those for ``flare1'' and ``flare2'' are slightly higher. This suggests a high probability that the plasma contributing to the SWCX originated from CMEs released during flares. While \citet{Reinard05} found a weak correlation ($r=0.18$) between oxygen ionization temperature and flare intensity, our current dataset consisting of only three data points is too small to identify any clear correlation in these SWCX observations.

Regarding Ne, Mg, Si, and Fe, a tendency for the ionization temperature ($T_{\rm{ion}}$) to increase with higher flare intensity was confirmed. This result suggests that coronal heating may proceed more efficiently in stronger flares, which is consistent with the positive correlation between flare intensity and plasma electron temperature ($T_{\rm{e}}$) reported by \citet{Kahler+20}. Although the $T_{\rm{ion}}$ values derived from our SWCX observations are approximately 2--10 times higher than those of quiescent solar wind ions (\cite{von Steiger+11}), they remain systematically lower than typical flare electron temperatures ($T_{\rm{ion}} \lesssim 0.5 T_{\rm{e}}$; e.g., \cite{Yuda+97}; \cite{Kahler+20}). These findings suggest that while ions are heated to higher temperatures by flares compared to quiescent times, their ionization states freeze in before reaching ionization equilibrium. This implies that the timescale for CME ejection, $\tau_{\rm{cme}}$, is shorter than the timescale to reach ionization equilibrium, where $\tau_{\rm{ion}} \sim 10^3$~s. This scenario is consistent with MHD simulations assuming Petschek-type reconnection (\cite{Imada+11}).

\subsection{Freeze-in height}\label{ssec:62}
\begin{figure}[t]
 \centering
 \includegraphics[width=8cm]{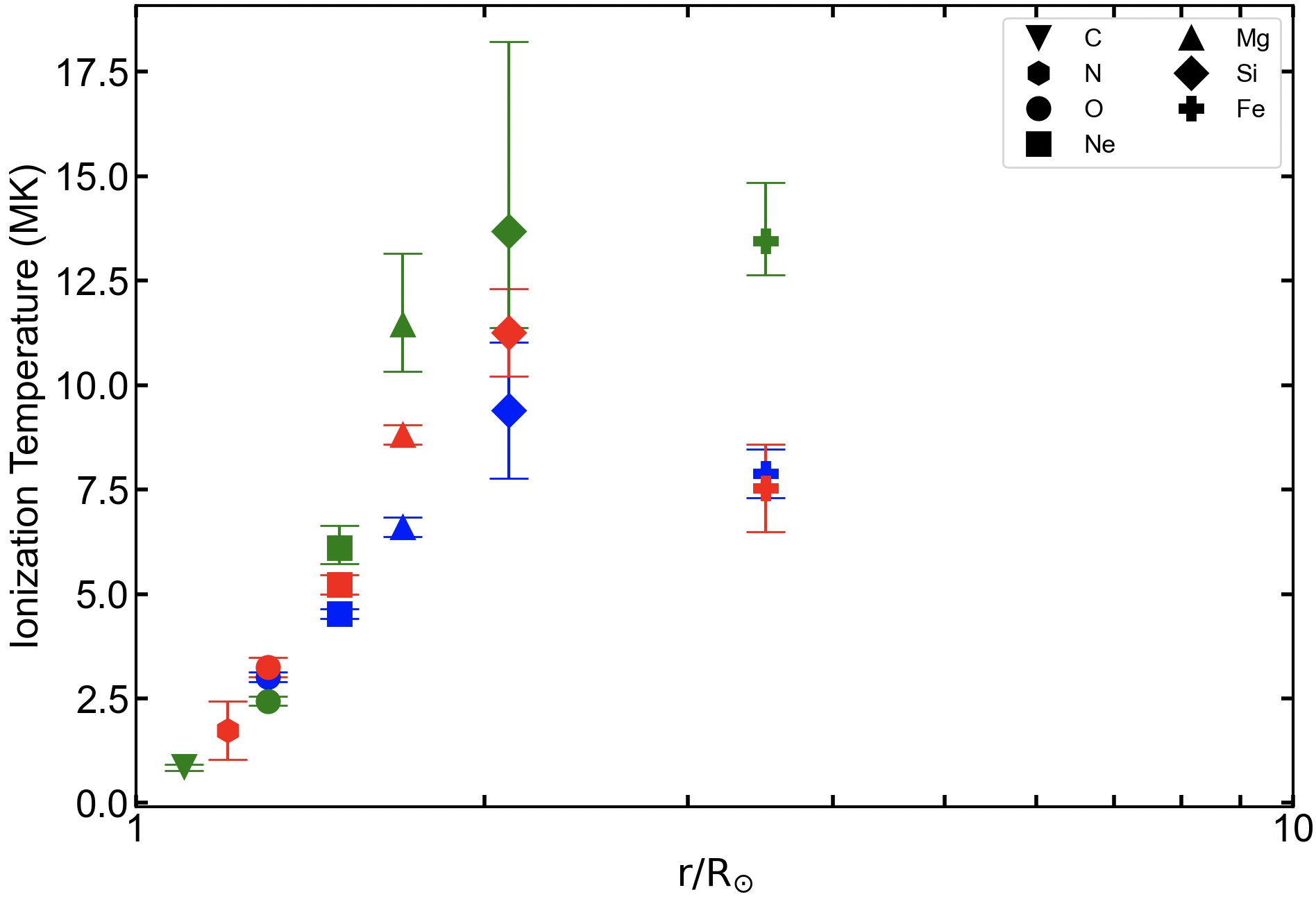}
 \caption{Schematic diagram showing the relationship between the freeze-in height and the ionization temperature obtained from the \texttt{vacx2} multi-temperature model fitting. Blue represents ``flare1'', red represents ``flare2'', and green represents \citet{Asakura+21}. The relative positions of the elements on the horizontal axis are based on \citet{von Steiger+11}. Note that while \citet{von Steiger+11} focuses on coronal holes, our observations target an active region, which is expected to have a different electron density profile. Because the freeze-in height strongly depends on the electron density, the absolute values for each element carry large uncertainties and are therefore omitted. The plot illustrates the qualitative trend that heavier elements generally freeze in at higher altitudes, with the ionization temperature reaching a maximum around the freeze-in height of Si. Alt text: A schematic scatter plot showing the relationship between freeze-in height and ionization temperature.
 }
 \label{fig:acx_temp_r}
\end{figure}

Another characteristic observed in the ionization temperature parameters is a tendency for heavier elements (elements with larger atomic numbers) to show higher ionization temperatures, with the exception of iron. A similar tendency has been reported in past observations of coronal holes (e.g., \cite{Geiss+95}; \cite{von Steiger+11}). This trend can be naturally explained by the elemental dependence of the freeze-in height.

The electron density at which freeze-in occurs differs for each element, and it is generally known that heavier elements freeze in at lower electron densities (e.g., \cite{Geiss+95}; \cite{Rakowski+11}). Assuming that the electron density decreases monotonically with height from the solar photosphere, the freeze-in height becomes higher for heavier elements. Based on \citet{Geiss+95} and \citet{von Steiger+11}, we plot the freeze-in height of each element on the horizontal axis and the ionization temperature parameters derived from the SWCX observations on the vertical axis in Figure \ref{fig:acx_temp_r}. As expected in this scenario, lighter elements such as carbon, nitrogen and oxygen freeze in at relatively lower altitudes. Therefore, the time from the onset of ionization to freeze-in is short, and they can be interpreted as being released while still in a low ionization state. On the other hand, neon, magnesium and silicon freeze in at higher altitudes, allowing more time for ionization, which suggests they were released after reaching a higher ionization state. 

However, the ionization temperature of iron, the heaviest element among them, is lower than that of silicon. At least two factors can be considered for this point. First is the possibility that the freeze-in height of iron was located at a higher altitude than the heating region. In this case, iron might have reached its maximum ionization temperature at a certain altitude, but then recombined at higher altitudes, resulting in a lower degree of ionization. This picture is consistent with models where the coronal electron temperature has a maximum value at a certain altitude (\cite{Geiss+95}).

Second is the possibility that the time required for iron ionization is significantly longer compared to other elements. It is known that for ionization by electron impact, the cross-section generally becomes smaller for heavier elements, reducing ionization efficiency (\cite{Bernshtam+00}). Although iron has a high freeze-in height and thus a long duration for ionization to proceed, its low ionization rate could result in it being frozen in at a lower degree of ionization than silicon.

Based on the above, for elements lighter than Si, the duration of the ionization process mainly determines the ionization state, resulting in higher degrees of ionization for heavier elements. On the other hand, for elements heavier than Si, the decrease in ionization efficiency becomes dominant, leading to a decrease in the degree of ionization as the atomic number increases. Since Si is the element where the balance between these two effects is optimal, considering that the ionization temperature of Si became the maximum allows for a consistent interpretation of our observational results.

\subsection{Elemental abundances}\label{ssec:63}
\begin{figure}[t]
 \centering
 \includegraphics[width=8cm]{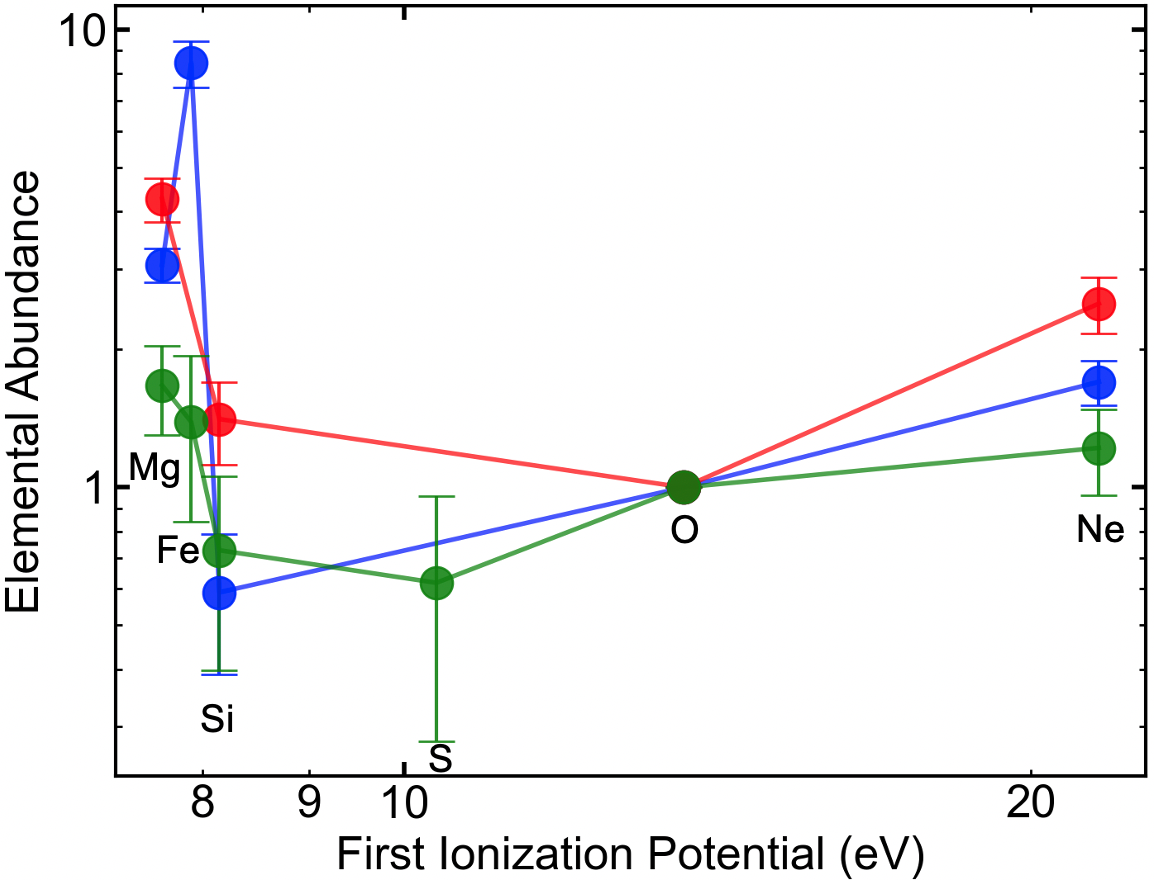}
 \caption{
  Relationship between the elemental abundance obtained from the \texttt{vacx2} multi-temperature model fitting and the First Ionization Potential (FIP). Abundances are shown as $(\mathrm{X}/\mathrm{O})/(\mathrm{X}/\mathrm{O})_\odot$, that is, relative to the photospheric composition and normalized to oxygen.} Blue represents ``flare1'', red represents ``flare2'', and green represents \citet{Asakura+21}. Alt text: A line graph displaying the relationship between elemental abundance and First Ionization Potential.
 
 \label{fig:fip}
\end{figure}
Figure \ref{fig:fip} shows the relationship between the elemental abundance obtained from the \texttt{vacx2} multi-temperature model fitting and the first ionization potential (FIP). The abundances are relative to the photospheric composition and are all normalized to oxygen. The characteristics of the inverse FIP effect, where the abundances of Si and S are reduced, can be confirmed. This shows a similar trend to the abundance patterns derived from the solar X-ray component reflected by the Earth's atmosphere (\cite{Katsuda+20}; \cite{Suzuki+25a}). A qualitatively similar tendency has also been reported from direct observations of solar flares. \citet{Dennis+15} found only modest enhancements for Si and S relative to the photospheric values. This behavior is consistent with the inverse FIP pattern inferred in this work. The ponderomotive-force model (\cite{Laming04,Laming12,Laming15,Laming21}), in which Alfvén or fast-mode waves can enhance or deplete low-FIP elements, provides a possible interpretation of this abundance pattern.

However, we note an apparent discrepancy in the Fe/Si abundance ratio between our result ($14.3 \pm 5.1$ for the ``flare1'' period, M5 class) and the value inferred from \citet{Suzuki+25a} ($0.65 \pm 0.13$ for M1--5 class flares). We speculate that this difference could be attributed to the multi-temperature nature of the Fe component in the source plasma. Specifically, the reflection spectra analyzed by \citet{Suzuki+25a} are considered to predominantly trace the high-temperature component ($\sim 1.5$~keV) located at the base of the corona. In contrast, our SWCX spectra are expected to probe the lower-temperature component ($\sim 0.7$~keV) that was ejected into interplanetary space as CMEs. Future statistical studies with a larger sample of SWCX events will be needed to fully verify this multi-temperature scenario.

Since SWCX emission is produced by the reaction of solar wind ions with neutral atoms after freeze-in, it is considered to reflect the elemental composition in the corona during flare occurrences relatively directly. These results suggest that SWCX observation is an effective method for investigating coronal abundances during flares, for which in-situ observations are difficult.

\section{Summary and conclusion}\label{sec:7}
In this paper, we analyzed SWCX emission in the Earth's exosphere using data obtained by XRISM/Xtend from 2024 August 9 to 13. We defined the periods before and during the soft X-ray enhancement as the ``stable'' and ``flare'' periods, respectively. Spectral analysis of the ``flare'' period revealed the detection of a rich variety of emission lines from C, N, O, Ne, and Fe, as well as heavy elements such as Mg and Si. Although the detection of these lines was previously limited by low photon statistics, the high sensitivity of Xtend enabled their detection with high significance ($> 15\sigma$).

Furthermore, based on the CME arrival times predicted by NASA CCMC, we extracted two specific intervals, ``flare1'' and ``flare2,'' from the ``flare'' period. Our subsequent spectral analysis revealed that the ion compositions of the plasmas were distinct between these periods. This difference in composition, combined with the consistency with the predicted arrival timing, leads to the conclusion that the enhancements in ``flare1'' and ``flare2'' were caused by two distinct CME events originating from different solar flares. This provides a first example in SWCX studies of separating multiple components of CME plasma within a single observation based on temporal spectral variations.

We performed spectral fitting using the AtomDB Charge Exchange model to determine the ionization temperatures and elemental abundances for each element. The key findings are summarized as follows:

\begin{itemize}
 \item A trend was observed where higher solar flare intensity corresponds to higher ionization temperatures. This suggests that coronal heating becomes more active as the flare intensity increases. This is the first such indication obtained from SWCX observations.
 \item The obtained ionization temperatures for each ion remained below 50\% of typical electron temperatures during flares. This indicates that the CME plasma was ejected into interplanetary space while still in a non-equilibrium ionization state. In this interpretation, the timescale for CME ejection is shorter than that required to reach ionization equilibrium, consistent with the scenario predicted by \citet{Imada+11}.
 \item With the exception of iron, a trend was confirmed where heavier elements (higher atomic numbers) exhibit higher ionization temperatures. This is likely attributable to the elemental dependence of the freeze-in height and heating efficiency.
 \item The derived elemental abundances showed a depletion of Si and S relative to O, confirming the characteristics of the Inverse FIP effect.
\end{itemize}

In conclusion, the large field of view and low instrumental background of XRISM/Xtend enabled us to obtain SWCX spectra of CME plasma with high photon statistics, including emission lines from heavy elements such as Mg and Si that are currently difficult to detect with in-situ observations such as those by ACE. These spectra allowed us to detect a rich variety of SWCX emission lines and to separate multiple components of CME plasma within a single observation based on temporal spectral variations. To our knowledge, this is the first study to use SWCX observations to discuss enhanced coronal heating with increasing flare intensity, CME ejection before the plasma reaches ionization equilibrium, and the elemental dependence of the freeze-in height. The abundance pattern shows depletion of Si and S relative to O. This is the second example of an inverse FIP pattern in SWCX observations after the event reported by \citet{Asakura+21}. These results confirm that SWCX observations can provide a detailed diagnostic of the ionization states and elemental abundances of coronal plasma during flares.




\begin{ack}
The authors thank Prof.\ Shinsuke Imada for valuable discussions. We also thank Dr.\ Yoshiaki Kanemaru for helpful advice on the evaluation of the Xtend vignetting profile and NXB. We also express our sincere gratitude to the XRISM/SWCX target team for their continuous support. This research has made use of data obtained from the XRISM mission, which is a collaborative mission between JAXA and NASA, with participation by ESA.
\end{ack}

\section*{Funding}
This work was supported by the JSPS Core-to-Core Program (Grant Number: JPJSCCA20220002), JSPS KAKENHI Grant Numbers 21H04972 and 26K07162, and the Tokyo Metropolitan University Emergent Future Society Research Project.

\section*{Data availability} 
The XRISM data analyzed in this study can be accessed through the HEASARC archive maintained by NASA/GSFC and the DARTS archive operated by JAXA/ISAS. The software used for the data reduction is publicly available through the HEASoft package.


\appendix 

\section*{Vignetting Profile}\label{app:vignetting}

In Section~\ref{sec:3}, we showed that the image during the ``flare'' period exhibits an enhancement over a broad region of the field of view (Figure~\ref{fig:image_comparison}). To quantitatively examine whether this spatial distribution is consistent with diffuse X-ray emission entering through the telescope optics, we calculated the radial profile of the 0.3--3.0~keV image obtained by subtracting the ``stable'' image from the ``flare'' image.

The current CALDB vignetting profile is based on a one-dimensional analytical approximation to measurements obtained during ground calibration. For Xtend, the approximation differs from the measurements by at least 10\% at off-axis angles greater than approximately 15~arcmin for the energies examined in the calibration.\footnote{\sloppy\url{https://heasarc.gsfc.nasa.gov/docs/xrism/calib/calib_docs/xrism_xma_caldb_vigncoef_v20240916.pdf}} Since our analysis extends to 25~arcmin from the optical axis, we used a 0.3--3.0~keV day-Earth image obtained from the XRISM trend archive in 2024 August as an empirical reference for the vignetting profile, following the approach of \citet{Suzuki+25b}. Since the day-Earth emission is bright and diffuse, its radial profile provides an empirical representation of the instrumental vignetting while reflecting the spatial response of the actual observation. We divided both images into annular regions with a width of 0.1~arcmin centered on the Xtend optical axis. For each annulus, we calculated the surface brightness after excluding the regions around point sources.

We fitted the radial profile of the difference image with a model of the form
\[
F(r) = A V(r),
\]
where \(V(r)\) is the empirical vignetting profile derived from the day-Earth image and \(A\) is a free normalization parameter. To minimize the effects of the regions excluded around point sources near the optical axis, we used the radial range of 2.5--25~arcmin for the fit.

\begin{figure}[t]
 \centering
 \includegraphics[width=8cm]{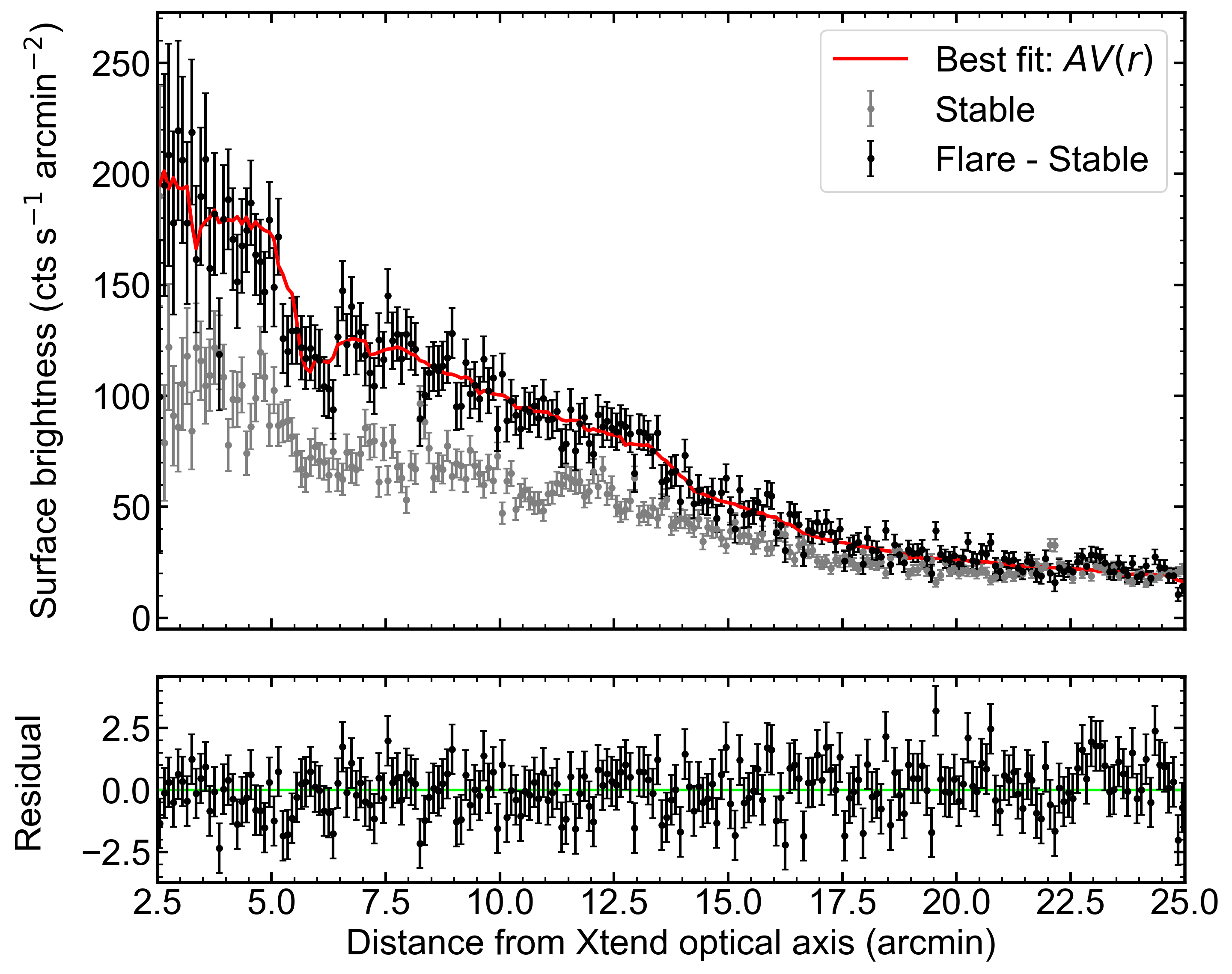}
 \caption{
 Radial profile of the background-subtracted 0.3--3.0~keV image around the Xtend optical axis. In the upper panel, the black points show the radial profile obtained by subtracting the ``stable'' image from the ``flare'' image. The gray points show the radial profile during the ``stable'' period for reference. The red curve shows the best-fit model \(A V(r)\), where \(V(r)\) is the empirical vignetting profile derived from the 0.3--3.0~keV day-Earth image. The lower panel shows the residuals from the best-fit model. The fit was performed in the radial range of 2.5--25~arcmin. Alt text: Radial profiles around the Xtend optical axis. The upper panel shows plotted points for the ``flare'' minus ``stable'' profile and the ``stable'' profile, together with a line graph of the best-fit day-Earth vignetting profile. The lower panel shows the residuals.
 }
 \label{fig:vigcurve}
\end{figure}

Figure~\ref{fig:vigcurve} shows the result. The normalization obtained from the fit was
\[
A = 195.085 \pm 1.584,
\]
with \(\chi^2/{\rm d.o.f.}=219.58/224\). The radial profile of the difference image is well described by the empirical vignetting profile without an additional constant component. This result supports the interpretation that the enhancement is dominated by diffuse X-ray emission entering through the telescope optics, rather than by localized emission from point sources or a spatially uniform residual background.

For reference, Figure~\ref{fig:vigcurve} also shows the radial profile during the ``stable'' period. The excess emission represented by the difference between the ``flare'' and ``stable'' images extends over a broad radial range rather than being confined to a localized region, which is consistent with diffuse emission.

\end{document}